\documentclass[usenatbib,useAMS, usegraphicx]{mn2e}

\usepackage{graphics,psfig,graphicx,bm}
\begin{document}

\title{Calculated spectra for HeH$^+$ and its effect on the opacity of cool metal poor stars}

\author[Elodie A. Engel, Natasha Doss, Gregory J. Harris and Jonathan Tennyson]{Elodie A. Engel, Natasha Doss, Gregory J. Harris and Jonathan Tennyson \\
Department of Physics and Astronomy, University College London,
London, WC1E 6BT}
\maketitle

\begin{abstract}
The wavelength and Einstein A coefficient are calculated for all
rotation-vibration transitions of $^4$He$^1$H$^+$, $^3$He$^1$H$^+$,
$^4$He$^2$H$^+$ and $^3$He$^2$H$^+$, giving a complete line list and
the partition function for $^4$HeH$^+$ and its isotopologues. This
opacity is included in the calculation of the total opacity of
low-metallicity stars and its effect is analysed for different
conditions of temperature, density and hydrogen number fraction. For a
low helium number fraction (as in the Sun), it is found that HeH$^+$
has a visible but small effect for very low densities ($\rho\leq
10^{-10}$g cm$^{-3}$), at temperatures around 3500 K. However, for
high helium number fraction, the effect of HeH$^+$ becomes important
for higher densities ($\rho\leq 10^{-6}$g cm$^{-3}$), its effect being
most important for a temperature around 3500 K. Synthetic spectra for
a variety of different conditions are presented.
\end{abstract}

\section{INTRODUCTION}

HeH$^+$ is thought to be the first molecular species to appear in the
Universe (Lepp et al. 2002). Consequently, stars formed from
primordial material should contain some HeH$^+$. Such stars probably
exist, Kashlinsky and Rees (1983) proposed a route by which stars of
mass as low as 0.2 M$_\odot$ could be formed from primordial
material. Moreover, the recent discovery of the very metal poor star
HE 0107-5240 (Christlieb et al. 2002) has added credence to the
possibility that low-mass primordial stars were formed. Observing such
stars would provide much information about the early Universe, and
could give an estimate of its age. Then, HeH$^+$ could influence their
formation and evolution. To have accurate information about the
stellar formation and evolution, we need to have the necessary opacity
data.

For zero-metallicity stars, the only contributors to the opacity are
the various hydrogen and helium species, ions and electrons. Many
hydrogen and helium species have already been included in calculated
opacity of zero-metallicity stars but HeH$^+$ has always been
neglected in such calculations (Harris et al. 2003). However, HeH$^+$
could be important because of its strong electric dipole moment. There
appears to be no available opacity data for HeH$^+$ and its
isotopologues. Such an opacity is calculated in this work. These
calculations provide predictions of the line intensity of HeH$^+$,
which also could be used to search for this molecular ion in the
Universe.

There have been several previous attempts to detect extraterrestrial
HeH$^+$: none conclusive. Moorhead et al. (1987) tried to detect
HeH$^+$ in the planetary nebula NGC 7027 but failed. According to
Cecchi-Pestellini and Dalgarno (1993), this failure arose from an
incorrect estimate of the chemistry of HeH$^+$ and hence of its
concentration in the planetary nebula. Furthermore Moorhead et
al. (1987) may not have observed the right volume of the
nebula. Miller et al. (1992) tentatively identified two emission
features in the spectra of supernova 1987A as due to HeH$^+$. This
observation has not been confirmed. The HeH$^+$ j=1-0 rotational
emission line may have been observed by the Infrared Space Observatory
in NGC 7027 (Liu et al. 1997). Its identification is complicated by
the near coincidence of its 149.14 $\mu$m rotational transition with
the 149.09 and 149.39 $\mu$m lines of CH. Moreover, Rabadan et
al. (1998) found that the electron impact
excitation rate to the $j=2$ rotational level from $j=0$ is
half that for $j=1-0$. As a result, observation of both $j=2-1$ and
$j=1-0$ emission lines should be possible. Observation of both these
would give a much more
secure detection of HeH$^+$.

Having accurate theoretical predictions for the spectrum of
low-metallicity stars should help to determine the best conditions to
search for and detect HeH$^+$.  Knowing the spectrum of HeH$^+$ should
also aid its detection in other astrophysical environments. Zygelman
et al. (1998) investigated the possibility for enhancement of the rate
of formation of HeH$^+$ in astrophysical environments when stimulated
by the cosmic background radiation field. They found that the effects
on the fractional abundance of HeH$^+$ are small in the early
Universe, in supernova ejecta and in planetary nebulae but may be
important in quasar broad-line clouds.

As our calculations show (see below), it could be possible to detect
HeH$^+$ in cool helium stars. Such stars exist, Saio and Jeffery
(2000) have observed and studied helium stars, and particularly
V625 Her, for more than 20 years. They showed that the merger of two
white dwarfs could lead to an extreme helium star (containing a number
fraction of helium around 0.9). On the other hand, Bergeron and
Leggett (2002) studied two cool white dwarfs SDSS 1337+00 and LHS 3250
and found that their observed energy distributions were  best
reproduced by models of cool extreme helium stars, and so these
stars should be such objects. However, they did not take account of
HeH$^+$ in their atmosphere models (Bergeron et al. 2001). Similarly
Stancil (1994) investigated the effect of the opacities of He$_2^+$
and H$_2^+$ in cool white dwarfs but neglected the effect of HeH$^+$
which is however a more stable ion.

In this work we will investigate the effect of HeH$^+$ on the opacity
of zero-metallicity stars. In section 2 we will describe the steps
leading to the calculation of the energy levels of all stable
isotopologues of HeH$^+$ and compute a spectroscopic line list and
their partition function. In section 3 we explain how the opacity
functions were calculated and use them in models of metal poor stars
for different temperatures, densities and hydrogen number
fractions. Finally, results are discussed in section 4.

\section{CALCULATION OF THE LINE LIST}
\subsection{Transition Data}

The program LEVEL 7.5 (Le Roy 2002) was used to obtain
rotation-vibration transition frequencies and Einstein A
coefficients. The core of the program determines
discrete eigenvalues and eigenfunctions of the radial
Schroedinger equation. We used LEVEL to calculate
the number, energy and wavefunction of each
vibration-rotational level for a given one-dimensional potential. These
were then used to
calculate Einstein $A$ coefficients for dipole transitions
between all bound vibration-rotation levels of each isotopologue.

The Schroedinger equation which is numerically solved by the program is :

\begin{equation}
	-\frac{\hbar^2}{2\mu}\frac{d^2\psi_{v,J}(R)}{dR^2}+V_J(R)\psi_{v,J}(R)=E_{v,J}\psi_{v,J}(R)
\end{equation}
where $\mu$ is the effective reduced mass of the system, $J$ the
rotational quantum number, $v$ the vibrational quantum number,
$V_J(R)$ is the effective potential of the molecule, $R$ being the
distance between the two atoms of the molecule. $V_J(R)$ is defined
by:

\begin{equation}
	V_J(R)= V_{BO}(R)+\frac{J(J+1)}{2\mu R^2}+\frac{V_{ad}(R)}{\mu}
\end{equation}
where $V_{BO}$ is the potential in the Born Oppenheimer approximation
and $V_{ad}$ are the adiabatic corrections.

The core of the calculation is to determine the eigenvalues $E_{v,J}$
and eigenfunctions $\psi_{v,J}(R)$ of the effective potential
$V_J(R)$. In order to obtain such information, one has to provide
input data to the program. The physical data are the masses of the
atoms of the diatomic molecule, its effective potential as a function
of the internuclear distance $R$ and the dipole moment, also as a
function of the internuclear distance $R$.

For each isotopologue, $V_{BO}$ was that  calculated using
the Born-Oppenheimer approximation by Kolos and Peek (1976). This
potential was corrected for adiabatic effects (Bishop and Cheung
1979), which accounts for some of the coupling between electronic and
nuclear motion. The corrections were given by Bishop and Cheung (1979)
for $^4$HeH$^+$ but as $^4$HeH$^+$ and its isotopologues $^4$HeD$^+$,
$^3$HeH$^+$ and $^3$HeD$^+$ only differ in their reduced mass, we just
had to replace the $^4$HeH$^+$ reduced mass by the appropriate
isotopologue's reduced mass.

The effective potential allows for adiabatic effects. Adjusting the
masses allows, at least partially, for non-adiabatic effects. We
calculated energy levels for several reduced masses given by Coxon and
Hajigeorgiou (1999). We tested the nuclear reduced mass, $\mu_{nucl}$,
which does not take account of the electrons, the atomic reduced mass
$\mu_{at}$, which takes full account of the electrons and the
effective, $\mu_{eff}$, and dissociation, $\mu_{diss}$, reduced masses
which have intermediate values. We compared our transition frequencies
calculated using the dipole selection rules ($\Delta J$=$\pm$1,
$\Delta v$ any) with the experimental transition frequencies tabulated
by Coxon and Hajigeorgiou (1999). The results are summarised in tables
1 and 2. Use of the dissociation reduced mass leads to the best
results (see tables 1 and 2). Since the dissociation reduced mass is
the most appropriate for high-lying (near dissociation) levels, it
should also be the best for levels yet to be observed
experimentally. We therefore used this mass for all further
calculations presented here.

\begin{table*}
{\begin{minipage}{140mm}
\caption{Difference between the experimental rotation-vibrational 
transition frequencies and the theoretical ones calculated with
different reduced masses for $^4$HeH$^+$, the experimental
frequencies and the reduced masses are from Coxon and Hajigeorgiou
(1999).}
\begin{tabular}{r r r r r r r r} \hline
& & & Experiment (cm$^{-1}$)
& \multicolumn{4}{|c}{Difference obs-calc (cm$^{-1}$) } \\ 
 v'& v''& J & P(J) transition & $\mu_{nucl}$ & $\mu_{at}$ & $\mu_{eff}$ & $\mu_{
diss}$  \\ \hline \hline
1 & 0 &  1 & 2843.9035 & 0.284   & $-$0.457 & $-$0.036 & 0.216 \\
  &   &  2 & 2771.8059 & 0.279   & $-$0.418 & $-$0.022 & 0.215 \\
  &   &  3 & 2695.0500 & 0.278   & $-$0.371 & $-$0.002 & 0.218 \\
  &   &  4 & 2614.0295 & 0.278   & $-$0.320 &  0.020   & 0.223 \\
  &   &  5 & 2529.134  & 0.278   & $-$0.266 &  0.043   & 0.228 \\
  &   &  6 & 2440.742  & 0.276   & $-$0.211 &  0.066   & 0.231 \\
  &   &  9 & 2158.140  & 0.268   & $-$0.037 &  0.136   & 0.240 \\
  &   & 10 & 2059.210  & 0.260   &  0.019   &  0.156   & 0.238 \\
  &   & 11 & 1958.388  & 0.250   &  0.074   &  0.174   & 0.234 \\
  &   & 12 & 1855.905  & 0.243   &  0.132   &  0.195   & 0.233 \\
  &   & 13 & 1751.971  & 0.225   &  0.181   &  0.206   & 0.221 \\ \hline       
2 & 1 &  1 & 2542.531  & 0.069   & $-$0.492 & $-$0.173 & 0.018 \\
  &   &  2 & 2475.814  & 0.067   & $-$0.456 & $-$0.159 & 0.019 \\
  &   &  5 & 2248.854  & 0.064   & $-$0.319 & $-$0.102 & 0.029 \\
  &   &  6 & 2165.485  & 0.070   & $-$0.260 & $-$0.072 & 0.040 \\
  &   &  7 & 2078.841  & 0.091   & $-$0.184 & $-$0.027 & 0.066 \\
  &   &  8 & 1989.251  & 0.089   & $-$0.129 & $-$0.005 & 0.069 \\
  &   &  9 & 1896.992  & 0.106   & $-$0.053 &  0.038   & 0.092 \\
  &   & 10 & 1802.349  & 0.114   &  0.016   &  0.072   & 0.105 \\
  &   & 11 & 1705.543  & 0.121   &  0.085   &  0.105   & 0.117 \\
  &   & 19 &  862.529  & 0.106   &  0.651   &  0.341   & 0.156 \\
  &   & 20 &  745.624  & 0.099   &  0.744   &  0.377   & 0.158 \\ \hline       
3 & 2 &  5 & 1966.356  & 0.043   & $-$0.175 & $-$0.051 & 0.023 \\
  &   & 17 &  833.640  & 0.096   &  0.650   &  0.336   & 0.147 \\
  &   & 18 &  719.769  & 0.098   &  0.752   &  0.380   & 0.158 \\ \hline
5 & 4 & 11 &  901.963  & 0.060   &  0.522   &  0.259   & 0.102 \\ 
  &   & 12 &  807.806  & 0.078   &  0.614   &  0.309   & 0.127 \\ \hline
6 & 5 &  8 &  863.378  & 0.048   &  0.521   &  0.252   & 0.092 \\ 
  &   &  9 &  782.925  & 0.044   &  0.575   &  0.274   & 0.093 \\  \hline
7 & 6 &  4 &  817.337  & 0.127   &  0.624   &  0.342   & 0.173 \\
  &   &  5 &  760.367  & 0.151   &  0.681   &  0.380   & 0.200 \\ \hline
7 & 5 & 12 &  938.200  &$-$0.327 &  1.758   &  0.573   &$-$0.136 \\ \hline
\end{tabular} 
\end{minipage}}
\end{table*}

\begin{table*}
{\begin{minipage}{140mm}
\caption{Difference between the 
experimental pure rotational transition frequencies and the
theoretical ones calculated with different reduced masses for
$^4$HeH$^+$, the experimental results and the reduced masses are from
Coxon and Hajigeorgiou (1999).}
\begin{tabular}{r r r r r r r r} \hline
& & & Experiment (cm$^{-1}$)
& \multicolumn{4}{|c}{Difference obs-calc (cm$^{-1}$) } \\ 
 v & J''& J' & frequency & $\mu_{nucl}$ & $\mu_{at}$ & $\mu_{eff}$ & $\mu_{diss}$  \\ \hline \hline
0 &  0 &  1 & 67.053  & 0.002 & $-$0.037& $-$0.015 & $-$0.002\\
  &  1 &  2 & 133.717 & 0.004 & $-$0.074& $-$0.030 & $-$0.003\\
  &  6 &  7 & 448.160 & 0.007 & $-$0.244& $-$0.102 & $-$0.016\\
  & 10 & 11 & 657.221 & 0.026 & $-$0.313& $-$0.121 & $-$0.005\\
  & 11 & 12 & 701.317 & 0.020 & $-$0.333& $-$0.132 & $-$0.012\\
  & 12 & 13 & 741.706 & 0.026 & $-$0.336& $-$0.130 & $-$0.007\\
  & 13 & 14 & 778.224 & 0.032 & $-$0.335& $-$0.127 & $-$0.002\\
  & 14 & 15 & 810.708 & 0.038 & $-$0.330& $-$0.121 &  0.004\\
  & 15 & 16 & 839.010 & 0.042 & $-$0.322& $-$0.115 &  0.009\\
  & 16 & 17 & 862.984 & 0.043 & $-$0.313& $-$0.110 &  0.011\\
  & 17 & 18 & 882.475 & 0.047 & $-$0.297& $-$0.102 &  0.015\\
  & 18 & 19 & 897.334 & 0.035 & $-$0.290& $-$0.106 &  0.005\\
  & 20 & 21 & 912.242 & 0.036 & $-$0.234& $-$0.080 &  0.011\\
  & 21 & 22 & 911.704 & 0.033 & $-$0.200& $-$0.068 &  0.011\\
  & 23 & 24 & 891.888 & 0.035 & $-$0.087& $-$0.018 &  0.024\\
  & 24 & 25 & 870.298 & 0.028 & $-$0.009&  0.012   &  0.025\\
  & 25 & 26 & 837.180 & 0.033 &  0.133  &  0.076   &  0.042\\ \hline
1 & 10 & 11 & 598.829 & 0.018 & $-$0.270& $-$0.107 & $-$0.009 \\
  & 11 & 12 & 637.767 & 0.013 & $-$0.282& $-$0.114 & $-$0.014\\
  & 12 & 13 & 672.989 & 0.009 & $-$0.291& $-$0.120 & $-$0.019\\
  & 13 & 14 & 704.270 & 0.019 & $-$0.281& $-$0.111 & $-$0.009\\
  & 14 & 15 & 731.430 & 0.016 & $-$0.280& $-$0.112 & $-$0.012\\
  & 15 & 16 & 754.235 & 0.019 & $-$0.266& $-$0.104 & $-$0.007\\
  & 16 & 17 & 772.464 & 0.017 & $-$0.253& $-$0.100 & $-$0.008\\
  & 17 & 18 & 785.837 & 0.015 & $-$0.235& $-$0.093 & $-$0.008\\
  & 18 & 19 & 793.997 & 0.023 & $-$0.197& $-$0.072 &  0.003\\
  & 19 & 20 & 796.490 & 0.027 & $-$0.156& $-$0.052 &  0.010\\
  & 20 & 21 & 792.616 & 0.032 & $-$0.101& $-$0.025 &  0.020\\
  & 21 & 22 & 781.245 & 0.048 & $-$0.019&  0.019   &  0.042\\
  & 22 & 23 & 760.340 & 0.030 &  0.061  &  0.043   &  0.033\\
  & 23 & 24 & 724.933 & 0.043 &  0.254  &  0.135   &  0.063\\\hline 
2 & 13 & 14 & 627.320 & 0.019 & $-$0.207& $-$0.079 & $-$0.002\\ 
  & 14 & 15 & 648.324 & 0.014 & $-$0.200& $-$0.078 & $-$0.005\\
  & 15 & 16 & 664.559 & 0.015 & $-$0.180& $-$0.069 & $-$0.003\\
  & 16 & 17 & 675.609 & 0.012 & $-$0.156& $-$0.061 & $-$0.003\\
  & 17 & 18 & 680.895 & 0.014 & $-$0.119& $-$0.044 &  0.001\\
  & 18 & 19 & 679.586 & 0.019 & $-$0.064& $-$0.017 &  0.011\\
  & 19 & 20 & 670.340 & 0.024 &  0.010  &  0.018   &  0.023\\
  & 20 & 21 & 650.613 & 0.025 &  0.118  &  0.065   &  0.033 \\ \hline
\end{tabular}
\end{minipage}}
\end{table*}

Using SI units, the Einstein $A_{(v' J')\rightarrow(v'' J'')}$
coefficient is related to the transition dipole by:

\begin{equation}
	|\langle\psi_{v' J'}|\mu|\psi_{v'' J''}\rangle|^2=\frac{2J'+1}{(2J''+1)\nu^3}A_{(v' J')\rightarrow(v'' J'')}
\end{equation}



We used the electric dipole moment data for $^4$HeH$^+$ given in Saenz
(2003). These data could not be used directly as Saenz tabulates only
the electronic dipole moment which is the contribution of the
electrons to the dipole moment, without taking account of the nuclear
charges of H and He. To use these data we first had to take account of
H and He and then to translate the dipole to the centre of mass for
each isotopologue, our results are given in table 3.

\begin{table*}
{\begin{minipage}{140mm}
\caption{Dipole moment, in atomic units, as a function of internuclear
separation, $R$, for the electronic ground state of
$^4$HeH$^+$ and isotopologues. The electronic
contribution is taken from Saenz (2003).}
\begin{tabular}{l r r r r} \hline
$R$ / a$_0$& $^4$HeH$^+$ & $^3$HeH$^+$&$^4$HeD$^+$&$^3$HeD$^+$\\ 
\hline \hline
0.6	&0.14879	&0.11879	&0.06879	&0.02879\\
0.8	&0.24025	&0.20025	&0.13358	&0.08025\\
1.0	&0.35267	&0.30267	&0.21934	&0.15267\\
1.1	&0.41620	&0.36120	&0.26953	&0.19620\\
1.2	&0.48426	&0.42426	&0.32426	&0.24426\\
1.3	&0.55658	&0.49158	&0.38325	&0.29658\\
1.4	&0.63288	&0.56288	&0.44621	&0.35288\\
1.45	&0.67245	&0.59995	&0.47912	&0.38245\\
1.5	&0.71291	&0.63791	&0.51291	&0.41291\\
1.6	&0.79641	&0.71641	&0.58308	&0.47641\\
1.8	&0.97277	&0.88277	&0.73277	&0.61277\\
2.0	&1.15986	&1.05986	&0.89319	&0.75986\\
2.2	&1.35549	&1.24549	&1.06216	&0.91549\\
2.4     &1.55743	&1.43743	&1.23743	&1.07743\\ \hline
\end{tabular}
\end{minipage}}
\end{table*}

Thus we obtained a line list of $^4$HeH$^+$, $^3$HeH$^+$, $^4$HeD$^+$
and $^3$HeD$^+$ transitions. Table 4 gives part of the line list: the
astronomically most important transitions for $^4$HeH$^+$. The full
line list for this main isotopologue which contains 1431 lines, and
the minor variants are available in electronic form at the Centre de
Donn$\acute{e}$es astronomiques de Strasbourg (CDS) via
http://cdsweb.u-strasbg.fr/cgi-bin/qcat?/MNRAS/???/???. These tables
use the standard Kurucz format (Kurucz 2000).

\begin{table*}
{\begin{minipage}{140mm}
\caption{Part of the $^4$HeH$^+$ line list: the astronomical most important
 transitions. The format is due to Kurucz (1993). The complete list
 can be obtained in electronic form via
 http://cdsweb.u-strasbg.fr/cgi-bin/qcat?/MNRAS/???/???.}
\begin{tabular}{ r r r r r r r r r } \hline 
   $\lambda$(nm) & A (s$^{-1}$) & J'  &  E'(cm$^{-1}$) & J"  &  E"(cm$^{-1}$) &code V' &  V''&   iso\\
          &              &     &           &     &           &    label' & label''& \\ \hline \hline
    1770.0593 & 0.428986E+02 & 3.0 & 5850.293 & 2.0 &   200.765 &102X 2 &    X 0 &    14\\
    1780.5247 & 0.384090E+02 & 2.0 & 5683.373 & 1.0 &    67.051 &102X 2 &    X 0 &    14\\   
    1794.8580 & 0.305583E+02 & 1.0 & 5571.471 & 0.0 &     0.000 &102X 2 &    X 0 &    14\\ 
    1835.4404 & 0.824290E+02 & 0.0 & 5515.334 & 1.0 &    67.051 &102X 2 &    X 0 &    14\\  
    1861.9524 & 0.517642E+02 & 1.0 & 5571.471 & 2.0 &   200.765 &102X 2 &    X 0 &    14\\  
    3248.6434 & 0.365347E+03 & 3.0 & 3278.973 & 2.0 &   200.765 &102X 1 &    X 0 &    14\\ 
    3301.8692 & 0.341654E+03 & 2.0 & 3095.638 & 1.0 &    67.051 &102X 1 &    X 0 &    14\\
    3363.8480 & 0.283663E+03 & 1.0 & 2972.786 & 0.0 &     0.000 &102X 1 &    X 0 &    14\\ 
    3516.0271 & 0.830802E+03 & 0.0 & 2911.170 & 1.0 &    67.051 &102X 1 &    X 0 &    14\\  
    3607.4759 & 0.542764E+03 & 1.0 & 2972.786 & 2.0 &   200.765 &102X 1 &    X 0 &    14\\
   50098.8713 & 0.375132E+01 & 3.0 &  400.370 & 2.0 &   200.765 &102X 0 &    X 0 &    14\\
   74786.5782 & 0.104399E+01 & 2.0 &  200.765 & 1.0 &    67.051 &102X 0 &    X 0 &    14\\
  149140.4805 & 0.109163E+00 & 1.0 &   67.051 & 0.0 &    0.000 &102X 0 &    X 0 &    14\\ \hline
\end{tabular}
\end{minipage}}
\end{table*}

\subsection{The Partition Function}

The internal partition function is defined by:

\begin{equation}
	Q_{vr}=\sum_ig_i\exp\left(\frac{-E_i}{kT}\right)
\end{equation}
where $k$ is Boltzmann's constant. The energy of the $i^{th}$ level
relative to the vibration rotation ground state is given by $E_i$ and
its degeneracy by $g_i$. $g_i$ is given by $(2J+1)$.  As we computed
all the energy levels, we used this data list and made a direct
summation of each level i. The partition functions of $^4$HeH$^+$ and
its isotopologues are given in table 5.

\begin{table}
\caption{Partition functions of $^4$HeH$^+$, $^3$HeH$^+$, $^4$HeD$^+$ and $^3$HeD$^+$ calculated by direct summation and by Sauval and Tatum (1984).}
\begin{tabular}{|r|r|r|r|r|r|} \hline
 $T$ [K] & $^4$HeH$^+$ &  $^4$HeH$^+$$^a$ & $^3$HeH$^+$ & $^4$HeD$^+$ & $^3$HeD$^+$ \\ \hline \hline
     10 &    1.00 &        &   1.00 &    1.01 &    1.00   \\
     20 &    1.02 &	   &   1.02 &    1.16 &    1.12   \\
     50 &    1.45 &	   &   1.40 &    2.09 &    1.93   \\
    100 &    2.44 &	   &   2.32 &    3.78 &    3.45   \\
    200 &    4.51 &	   &   4.26 &    7.20 &    6.54   \\
    400 &    8.69 &	   &   8.19 &   14.11 &   12.77   \\
    600 &   12.93 &	   &  12.17 &   21.17 &   19.12   \\
    800 &   17.28 &	   &  16.24 &   28.59 &   25.76   \\
   1000 &   21.84 &  20.36 &  20.51 &   36.62 &   32.90   \\
   1500 &   34.79 &  32.36 &  32.53 &   60.43 &   53.87   \\
   2000 &   50.76 &  47.66 &  48.28 &   90.91 &   80.53   \\
   2500 &   70.54 &  66.55 &  65.52 &  129.45 &  114.10   \\
   3000 &   94.80 &  89.33 &  87.83 &  177.32 &  155.70   \\
   3500 &  123.90 & 116.35 & 114.58 &  235.36 &  206.00   \\
   4000 &  157.84 & 147.93 & 145.77 &  303.68 &  265.09   \\
   4500 &  196.32 & 184.42 & 181.10 &  381.76 &  332.47   \\
   5000 &  238.77 & 226.19 & 220.08 &  468.57 &  407.23   \\
   5500 &  284.56 & 273.61 & 262.10 &  562.82 &  488.24   \\
   6000 &  333.00 & 327.04 & 306.55 &  663.12 &  574.31   \\
   6500 &  383.44 & 386.87 & 352.82 &  768.13 &  664.27   \\
   7000 &  435.30 & 453.51 & 400.38 &  876.59 &  757.06   \\
   7500 &  488.05 & 527.35 & 448.75 &  987.41 &  851.74   \\
   8000 &  541.27 & 608.81 & 497.54 & 1099.61 &  947.50   \\
   8500 &  594.57 & 698.32 & 546.40 & 1212.41 & 1043.66   \\
   9000 &  647.67 & 796.29 & 595.07 & 1325.12 & 1139.65   \\ \hline
\multicolumn{6}{l}{$^a$ Sauval and Tatum (1984).}
\end{tabular}
\end{table}

We fitted the partition functions of the isotopologues to the standard
form of Irwin (1981):
\begin{equation}
	\log Q_{vr}= \sum_{n=0}^{n=3}a_n(\log(\theta))^n
\end{equation}
where $\theta$=$\frac{5040}{T}$. The coefficients a$_n$ are given in table 6.

\begin{table*}
{\begin{minipage}{140mm}   
\caption{Coefficients a$_n$ for the partition function of $^4$HeH$^+$, $^3$HeH$^
+$, $^4$HeD$^+$ and $^3$HeD$^+$.}
\begin{tabular}{ c c c c c c } \hline
 a$_n$ & $^4$HeH$^+$$^a$ & $^4$HeH$^+$ & $^3$HeH$^+$&$^4$HeD$^+$&$^3$HeD$^+$\\ 
\hline
   a$_0$&   2.3613 &    2.38224   &   2.34679  &   2.67542  &    2.61436   \\
   a$_1$&$-$1.9733 & $-$1.79357   & $-$1.78517  & $-$1.8795   & $-$1.86015 \\ 
   a$_2$&   0.6761 & $-$0.0539707 & $-$0.042402 & $-$0.108517 & $-$0.102426\\
   a$_3$&   0.0794 &    0.736706  &   0.726715 &   0.788349 &    0.782675  \\ 
\hline
\multicolumn{6}{l}{$^a$ Sauval and Tatum (1984).}
\end{tabular}
\end{minipage}}
\end{table*}

This fit reproduces all our data to within better than 0.9$\%$. For
$^4$HeH$^+$, our results agree reasonably with those of Sauval and
Tatum, although our $Q_{vr}$ is significantly lower at high
temperatures. We know of no data on the partition function for the
other isotopologues.

\section{The Opacity of Zero-Metallicity Stars}

\subsection{The Opacity of HeH$^+$}

The monochromatic opacity or absorption coefficient $\kappa_\nu$ is
given by the relation:

\begin{equation}
	\kappa_\nu=\sum_iS_if_i(\nu-\nu_{i0})
\end{equation}
where for the particular combination of units of dipole moment in
Debye, Einstein $A$ coefficients in s$^{-1}$ and wavenumber in
cm$^{-1}$, the integrated intensity $S_i$ in cm per molecule, is given
by:

\begin{eqnarray}
S_i&=&\frac{1.3271\times10^{-12}(2J'+1)}{Q_{vr}\nu^2}\exp\left(\frac{-E''}{kT}\right)
\nonumber \\
& &\left[1-\exp\left(\frac{-hc\nu}{kT}\right)\right]A_{(v' J')\rightarrow(v'' J'')}
\end{eqnarray}
$f_i$ is the line profile, which is assumed to result from the Doppler
effect, $\nu_{i0}$ is the central frequency of line $i$ and $\nu$ is
the frequency where the monochromatic opacity is calculated. To
calculate $\kappa_\nu$ we took account of the lines whose frequency
$\nu_{i0}$ was in the region [$\nu-\Delta\nu,\nu+\Delta\nu$]. The
opacity function for one species is the set of all the values
$\kappa_\nu(\nu)$. So to compute the opacity function of the molecular
ion HeH$^+$, we had to obtain $\kappa_\nu$ for all frequencies. We
calculated the opacity functions of all the isotopologues of HeH$^+$
and included them in the program LoMES (Harris et al. 2004, to be
submitted) to take account of them in the calculation of the total
opacity, see next section.  In figure 1 we give the opacity of
$^4$HeH$^+$. Inspecting this shows that we can already assume that
this molecular ion will only have an effect on the opacity for
wavenumbers under 5000 cm$^{-1}$.

\begin{figure}   
\centerline{\includegraphics[scale=0.3]{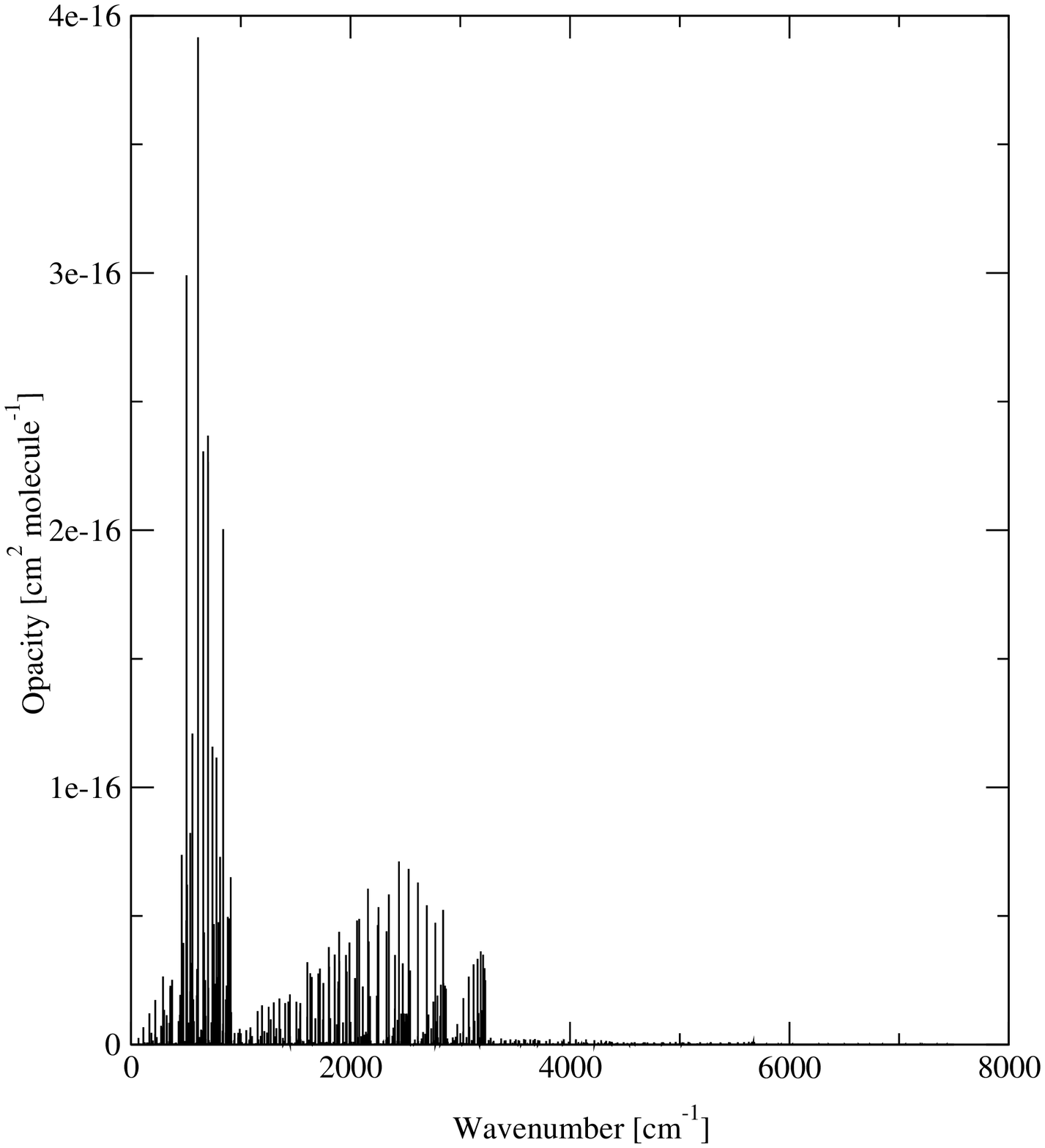}}
\caption{Opacity of $^4$HeH$^+$ at a temperature $T$=3500 K.}
\end{figure}
\vspace{0.5cm}

\subsection{Calculation of the Total Opacity} 

We used program LoMES which was written to calculate the effect of
H$_3^+$ (Harris et al. 2003). This program calculates the
frequency-dependent continuous opacity, which is the total opacity. It
uses the subroutines developed by Booth and Lynas-Gray (2002, private
communication) and Harris et al. (2003) to calculate the continuous
opacity contributions from H I, H$^-$, He$^-$, He I, He II bound-free
and free-free, H$_2^-$ free-free, Rayleigh scattering of He I, H$_2$
and H I, Thompson scattering by e$^-$ and H$_2^+$ free-free and
bound-free. Finally, H I line absorption are included by using the
STARK subroutine from the ATLAS12 program (Kurucz 1993). For more
details concerning program LoMES, see Harris et al. (2004, to be
submitted).$\\$ Harris et al. (2003) showed that H$_3^+$ influences
the opacity of metal poor stars. H$_3^+$ contributes to opacity both
through electron donation and by direct absorption. In fact, in a
zero-metallicity gas there are four different mechanisms which control
the opacity (Lenzuni et al. 1991): collision-induced absorption by
H$_2$; Rayleigh scattering by H, H$_2$ and He; Thomson scattering by
e$^-$ and free-free and bound-free absorption by H$^-$.$\\$ The last
mechanism dominates the opacity at densities which can be found in
stellar atmospheres (10$^{-10}$ g cm$^{-3}\leq\rho\leq 10^{-6}$ g
cm$^{-3}$), at temperatures between 3500 and 7000K and it was shown
(Harris et al. 2003) that H$_3^+$, acting as an electron donor,
increased the abundance of H$^-$ and hence the opacity. Furthermore,
H$_3^+$ was found to contribute up to 15$\%$ to the opacity via line
absorption. This is far smaller than its indirect effect on the
opacity through electron donation (it can increase the opacity by a
factor of three). On the other hand, the effect of H$_3^+$ on the
opacity is only important at densities higher than 10$^{-8}$ g
cm$^{-3}$. So, for very low-metallicity stars having a very low
density, the molecular ion HeH$^+$ could have an effect on their
opacity because of its strong dipole moment, particularly in cases
where the He to H ratio is high.

\subsubsection{The Number Density of HeH$^+$ and Its Isotopologues}

In this work we use the LoMES equation of state subroutine which is an
improved version of the equation of state used by Harris et
al. (2003). LoMES was developed for use in modelling of very
low-metallicity stellar atmospheres. It takes account of what are
likely to be the 12 most common elements in very low-metallicity
stars: H, He, C, N, O, Ne, Na, Mg, Si, S, Ca and Fe. The cat and
anions of these atoms are accounted for as are the molecular ions
H$_2^+$, H$_2^-$, H$_3^+$ and HeH$^+$. Finally LoMES currently
considers for the molecules: CO, CN, C$_2$, CH, OH, NH, NO, O$_2$,
N$_2$, CS, HCN, H$_2$O and their cat and anions.

LoMES uses the Saha equation to build a set of 13 non-linear
simultaneous conservation equations. These equations are then solved
for molecular equilibrium by using a multi-variable Newton-Raphson
iteration. This technique has been well documented in the past
(e.g. Kurucz 1970). The LoMES subroutine assumes a preset metal mix,
and takes input of hydrogen, helium and metal number fraction, as well
as temperature. The user can choose either density or pressure as the
final state variable. LoMES returns the density or pressure along with
the number densities of the 87 species including electrons.

In this work we focus on pure hydrogen-helium mixes, so that we set
the metal number fraction to near zero (10$^{-30}$). This allows us to
avoid the potential numerical problems of setting the number fraction
to zero, whilst the effects of the metals are insignificant.

The number density N$_{HeH}$ of HeH$^+$ was calculated by the program
and we had to find the proportions of the different isotopologues in
the number density of HeH$^+$. As we were interested in metal poor
stars formed from primordial material, we considered that the
proportions of the isotopologues of HeH$^+$ were the same as in the
primordial Universe.

To find these proportions, we used data given in Coc et
al. (2004). They used both standard big bang nucleosynthesis (SBBN)
and recent WMAP results to calculate the primordial $^4$He mass
fraction and the abundance ratio of D/H and $^3$He/H. Both SBBN and
WMAP are used to determine the baryonic density (density of ordinary
matter) in the Universe. The SBBN is a method, based on nuclear
physics in the early Universe whereas WMAP studies the cosmic
microwave background anisotropies to deduce the baryonic parameter
$\Omega_bh^2$ (where $h$ is the Hubble constant expressed in units of
100 km s$^{-1}$ Mpc$^{-1}$). The WMAP+SBBN results were Y$_p$ = 0.2479
$\pm$ 0.0004 for the $^4$He mass fraction, $^3$He/H = (1.04 $\pm$
0.04) ×10$^{-5}$ and D/H = (2.60$^{+0.19}_{-0.17}$) ×10$^{-5}$.

Thus, we used number densities of :$\\$
1.25$\times$10$^{-4}$  N$_{HeH}$ for $^3$HeH$^+$,$\\$
2.60$\times$10$^{-5}$ N$_{HeH}$ for $^4$HeD$^+$,$\\$
3.26$\times$10$^{-9}$ N$_{HeH}$ for $^3$HeD$^+$,$\\$
[1- (1.25$\times$10$^{-4}$+2.60$\times$10$^{-5}$ +3.26$\times$10$^{-9}$)]N$_{HeH}$ for $^4$HeH$^+$.$\\$

The total opacity was obtained by summing each opacity multiplied by
the corresponding density fraction. We investigated the effect of
HeH$^+$ on the opacity of zero-metallicity stars for different
conditions, the results are discussed in the next section.

\begin{figure}
\centerline{\includegraphics[scale=0.3]{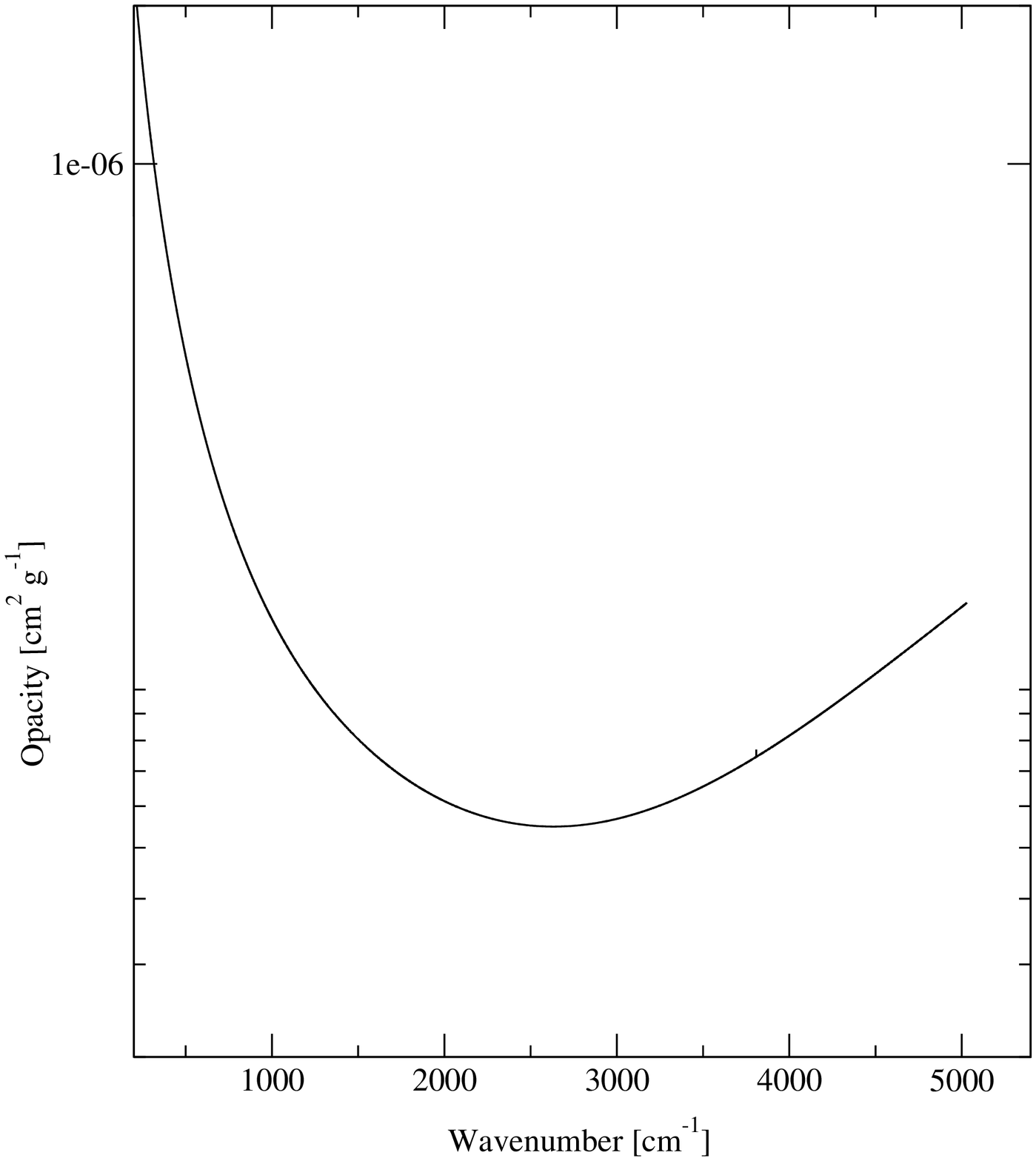}}
\centerline{\includegraphics[scale=0.3]{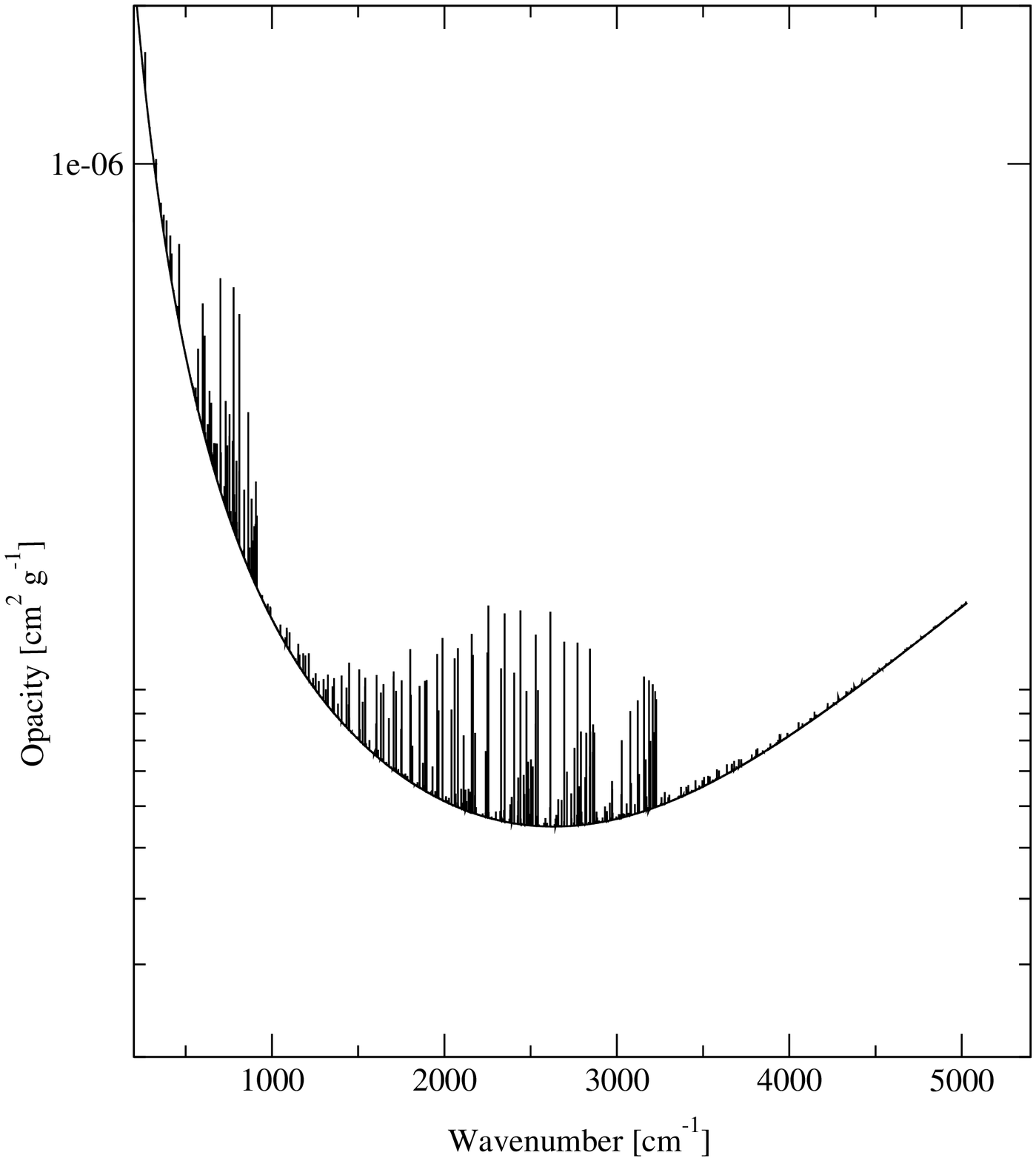}}
\caption{Opacity function taking account of all the hydrogen and 
helium species, including H$_3^+$ in the upper graph and including
H$_3^+$ and HeH$^+$ in the lower graph, for a density $\rho$=10$^{-9}$
g cm$^{-3}$, a temperature $T$=3500 K and a hydrogen number fraction
X$_n$=0.1.}
\end{figure}
\vspace{0.5cm}

\section{RESULTS AND DISCUSSION}

In order to find the importance of HeH$^+$ in zero-metallicity stars,
we have calculated the total opacity for different temperatures $T$,
densities $\rho$ and hydrogen number fractions X$_n$. The total
opacity takes account of all the hydrogen and helium species including
H$_3^+$ but with or without HeH$^+$ to investigate its effect.

First we made a calculation using the solar hydrogen number fraction
(X$_n$=0.92) and a temperature of 3500 K where the effect of H$_3^+$
is maximum (Harris et al. 2003). The effect of H$_3^+$ on the opacity
is only important at high densities ($\rho\geq$10$^{-8}$ g cm$^{-3}$).

As a result, for low-metallicity stars having very low density, we
find very weak lines caused by the molecular ion HeH$^+$. HeH$^+$
contributes to opacity only for wavenumbers between 0 and 5000
cm$^{-1}$ (see figure 1).

At temperatures higher than 4000 K neither HeH$^+$ lines nor H$^+_3$
lines are visible, that is because the opacity created by H$^-$
increases much faster with the temperature than the opacity created by
HeH$^+$ and the effect of H$^+_3$ is only important for temperatures
lower than 3500 K.

At lower temperatures, around 3000 K, H$^+_3$ lines dominate and no
HeH$^+$ lines are visible.

Finally, for densities higher then 10$^{-8}$ g cm$^{-3}$ one can
neglect HeH$^+$ lines if the number fraction of helium is
solar. However they become visible, but not very important, at
densities lower than 10$^{-10}$ g cm$^{-3}$ and for a temperature
around 3500 K.

We then decreased the hydrogen number fraction (X$_n$=0.8 to 0.1),
that means increased the helium number fraction. We found that HeH$^+$
lines become more important as the helium number fraction increases.

Keeping the hydrogen number fraction equal to 0.1, we varied the
temperature and the density. With this high helium number fraction,
HeH$^+$ lines are clearly visible for higher densities from 10$^{-8}$,
even 10$^{-6}$ g cm$^{-3}$, their maximum effect is for a density of
10$^{-9}$ g cm$^{-3}$ and a temperature of 3500 K (see figure 2).

For a high helium number fraction, HeH$^+$ lines become important even
for densities around 10$^{-8}$ g cm$^{-3}$ and are more important for
a temperature around 3500K. Thus the helium number fraction seems to
be the most important parameter in determining the role of
HeH$^+$. Thus, if one hopes to detect HeH$^+$ in a star's atmosphere,
the best conditions are a very low density ($\rho\leq 10^{-9}$g
cm$^{-3}$), a temperature around 3500 K and a high number fraction of
helium.

As H$_3^+$, HeH$^+$ influences the opacity by direct absorption and by
electron donation, but our tests showed the latter is unimportant
compared to the direct absorption. Indeed, we checked the effect of
HeH$^+$ as an electron donor. For a hydrogen number density of 0.1 or
less, the effect of the electron donation by HeH$^+$ becomes important
but only for high densities ($\rho\geq 10^{-4}$ g cm$^{-3}$).

\section{CONCLUSION}

We have calculated all the energy levels of $^4$HeH$^+$, $^3$HeH$^+$,
$^4$HeD$^+$ and $^3$HeD$^+$ and the Einstein $A$ coefficients
corresponding to each rotation-vibrational transitions and pure
rotational transitions. Thus we have obtained complete line lists for
all the isotopologues of HeH$^+$. These transition data could also
be important for nebular and interstellar studies.
The partition functions of each
isotopologues have been computed by direct summation of the energy
levels. These data have been then used to calculate the opacity of
HeH$^+$.

The total opacity has been calculated for zero-metallicity gas, taking
account of the isotopic composition of the primordial Universe. We
have used an updated version of the program which was used to
investigate the effect of H$_3^+$ by Harris et al. (2003). The total
opacity has been calculated for different temperatures, densities and
number fraction of hydrogen.

We have checked the validity of Harris et al.'s (2003) effective
neglect of the molecular ion HeH$^+$ when they studied the effect of
H$_3^+$. Their results are correct because they were interested in
high densities, around 10$^{-6}$ g cm$^{-3}$ and in a solar helium
number fraction. However, HeH$^+$ lines become visible for very low
densities (10$^{-10}$ g cm$^{-3}$) when the helium number fraction is
solar, and for temperatures near 3500 K. Moreover, when the helium
number fraction is high (0.9), HeH$^+$ lines can be strongly seen for
higher densities ($\rho\leq$10$^{-6}$ g cm$^{-3}$) and are still more
important for a temperature around 3500 K and a density of 10$^{-9}$ g
cm$^{-3}$.

The stars whose observed energy distributions were analysed by
Bergeron and Leggett (2002) seem to satisfy these conditions. Indeed,
Bergeron and Leggett (2002) studied two cool white dwarfs LHS 3250 and
SDSS 1337+00 and showed that the best predictions for their observed
energy distributions were given by simulations using extreme helium
compositions. They concluded that their cool white dwarfs should have
helium-rich atmospheres and effective temperatures below 4000 K. The
simulations closest to their observations for these stars were for low
masses (less than 0.7 M$_\odot$), temperatures around 3300 K and very
high He to H ratios (more than 10$^3$). These stars seem to be good
candidates to look for HeH$^+$. HeH$^+$ appears to have been largely
neglected in atmospheric models for cool white dwarfs (Stancil 1994,
Bergeron et al. 2001). A study on the effects of including HeH$^+$ in
cool helium rich white dwarfs will be presented elsewhere (Harris et
al. 2005).

\section*{ACKNOWLEDGEMENT}

We thank Tony Lynas-Gray for helpful discussions, the UK Particle
Physics and Astronomy Research Council (PPARC) and the Engineering and
Physical Sciences Research Council (EPSRC) for funding. EAE is also
grateful to the R$\acute{e}$gion Rh$\hat{o}$ne Alpes for supporting
her visit to University College London.

\bibliographystyle{/amp/tex/styles/MNRAS/mn2e}

\end{document}